# NREL Phase VI wind turbine in the dusty environment

J. Zare[1,2], S. E. Hosseini[1], M. R. Rastan[3,*]

[1] School of Mechanical Engineering, Iran University of Science and Technology, Tehran, Iran

[2] Iran Railways Company, Tehran, Iran

[3] Department of Engineering Mechanics, School of Naval Architecture, Ocean and Civil Engineering, Shanghai Jiao Tong University, 200240, Shanghai, China

*Corresponding author: rastan@sjtu.edu.cn

**Abstract**

The meteorological conditions markedly affect the energy efficiencies and cost/power rate of the wind turbines. This study numerically investigates the performance of the National Renewable Energy Laboratory (NREL) Phase VI wind turbine, designed to be insusceptible to the surface roughness, undergoing either clean or dusty air. First, the numerical approach is validated against the available experimental data for clean air. Following this, the model is developed into a Lagrangian-Eulerian multiphase approach to comprehensively analyze the effects of the dusty air. The dependence of aerodynamic performance on the wind speed ($U_\infty$ = 5-25 m/s), particle diameter ($d_p$ = 0.025-0.9 mm) and angle of attack ($AoA$ = 0°-44°) is investigated. It is found that the turbine performance generally deteriorates in dusty conditions. But it becomes relatively acute for $d_p \geq 0.1$ mm and post-stall state. As such, the generated power is reduced by 4.3% and 13.3% on average for the air with the $d_p$ = 0.05 and 0.9 mm, respectively. The particles change the flow field profoundly, declining the pressure difference between the suction/pressure sides of the blade-airfoil, advancing the boundary layer separation, and strengthening the recirculation zones. The above changes account for a lower lift coefficient and higher drag coefficient.

## 1. Introduction

Growing energy demands, environmental pollution concerns, and depletion of fossil fuel sources have escalated the interest in renewable energies [1]. Energy extraction through wind turbines is one of the first nominees for providing clean energy. 6.6% of the global electricity was generated by the wind in 2021, while it was 3.5% by 2015 [2], showing the crucial role of wind turbines in the world economy and energy demand. Because of the pivotal role of the horizontal axis turbines in industry [3], many endeavors have been devoted to studying the performance and optimizing this type, e.g. see [4-8].

The turbine output power is a function of various design factors [9-11], including blade twist angle, the motor rotational speed, etc. On the other hand, several studies showed significant roles of meteorological and environmental conditions on outputs [5, 12-19]. For instance, the numerical study of Wu et al. [5] illustrated how rain deteriorates the turbine performance; the larger the raindrop diameter, the more performance degradations (e.g. 24% lift coefficient reduction). As another common condition, especially for semi-arid regions like the Middle East, the dusty-wind environment has been subjected to a handful of studies [12-16]. The blade erosion and dust accumulation, i.e., perturbing the surface smoothness (especially at the leading edge near the stagnation point, see Fig 1*a*) by hitting the particles, are often simulated with rough blades (e.g. see [12-14]). It was experimentally shown that the power loss accretes with accumulating dust in time, e.g., ~50% after 270 days without cleaning blades of a 100kW stall-regulated wind turbine [13]. The alternative approach is pondering the output on a dusty farm (Fig. 2*b*) or focusing on the smooth blades' performance while facing the polluted wind (e.g. see [15]).

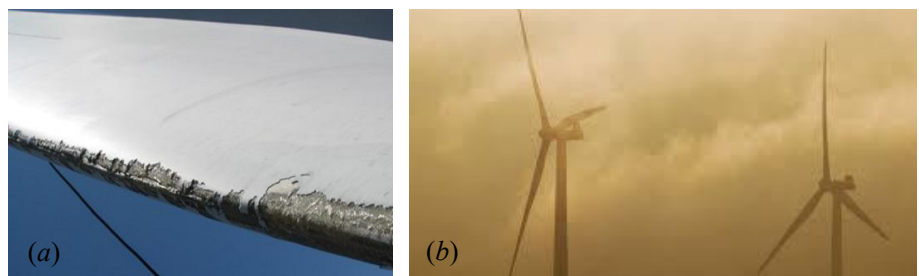

**Fig. 1**. (*a*) Rough surface of a wind turbine blade caused by dust [12] and (*b*) a farm under dusty wind[*].

[*] https://www.wind-watch.org/video-windfall.php

Diab et al. [12] explored dust accumulation on a number of widely used airfoil sections in the wind turbine industry by NACA, NREL and DU, showing how passing time changes the blade profile and correspondingly the aerodynamic parameters. The results show that although the dust accumulation percentage is contingent on the blade type, the time and adverse effects have almost a linear rate. They suggested installing a leading edge slat to mitigate the adverse influences of dust contamination. Khalfallah and Koliub [13] experimentally analyzed the effect of surface roughness due to dust accumulation on the performance of wind turbines. Moreover, the mechanism of dust accumulation was discussed. They observed that the roughness from the chord line towards the leading edge on blades of a stall-regulated, horizontal axis 300 kW wind turbine changes from 5% to 20%. Increasing the blade surface roughness generally reduces the effectiveness of the airfoil and the power output. However, the deterioration depends on the size and nature of the roughness, Reynolds number, and airfoil type. The 2D numerical study of Li et al. [14] on the DU 95-W-180 airfoil determined the critical roughness height of a wind turbine airfoil as 0.5 mm. The lift and drag coefficient curves of the airfoil with sub-critical roughness decreases and increases, respectively. Beyond the critical value, the rate of variations are decelerated.

Khakpour et al. [15] compared the pressure and aerodynamic force coefficients over a 2D section of an S819 airfoil under clean and dusty winds. They varied some parameters individually, such as the sand dimension and sand/air mass flow rate ratio, to get insights into the overall impact of airborne particles on the wind turbine performance. The particle size was determined as a critical parameter in judging the turbine performance in the dusty environment, although no data about the flow field or wake flow was provided. Issues concerning roughness on wind turbine blades were reviewed by Sagol et al. [16]. They discussed how different contaminations, such as dust, dirt, ice, and insects, lead to surface roughness and affect the flow field and power generation. It was emphasized that the turbine performance is a function of the size, location, and density of the elements causing roughness, especially accretion near the leading edge. However, it seems that the performance remains stable after a certain a level of roughness height.

The above literature review clarifies that studies on the dependence of wind turbine performance on dusty air are scarce. How the dust is accumulated on the blade, changes the flow pattern and degrades the performance has been focused in a few published studies. Generally, a clean oncoming air interacting a blade with distributed roughness was considered for these analyses. But, there is not enough data in the literature (except [15]) to show how the interaction of dusty wind (i.e., two-phase flow) and blades might change the turbine output. Comparing the roughness-considering viewpoint, the latter indicates how an already cleaned (or new) blade deal with the dusty wind or how the wind-induced power is correlated with the particle sizes. Besides, Wu et al. [5] recently compared 2D and 3D CFD approaches for simulating the rain dependence of wind turbine aerodynamic parameters and found discrepancies as large as 67%. The dust-induced performance and flow pattern due to interaction of multiphase flow and full-scale (or 3D) turbine have not been carried out so far. Specifically, this study aims to answer two questions: (i) what is the airborne particle mechanism (beyond the accumulation or roughness) of agitating the turbine flow pattern and degrading the engineering parameters? (ii) How those effects are correlated with the particle size? Or is there any possible safe zone (i.e., stable performance above/lower a certain size)? The answer may help to provide a suitable control mechanism to deal with dusty sites.

To address these knowledge gaps, first a three-dimensional CFD framework is validated against the available experimental data and employed to study the dusty environment influences on the operating parameters of a full-scale two-bladed horizontal axis wind turbine. Then, the clean airflow and dusty one with the airborne particles ($d_p$= 0.025-0.9 mm) are simulated over the wind turbine. A systematic comparison of the clean and dusty airflow is conducted for different wind speeds ($U_\infty$ = 5-25 m/s) and angles of attack ($AoA$ = 0°-44°), mostly through primary engineering parameters including aerodynamic force coefficients, surface pressure, wake velocity distribution, and boundary layer specifications. Next, the $AoA$ significance in dusty wind physics over turbine is investigated. Then, the wake flow and the blade boundary layer are scrutinized. Finally, to bridge the present study to those in which dust is as roughness, the variation of turbulent viscosity with $d_p$ is concisely investigated, as the primary parameter in dust accumulation.

## 2. Methodology

The NREL Phase VI wind turbine is selected in this study. The National Renewable Energy Laboratory (NREL) completed the analysis of the Phase VI wind turbine in the NASA Ames in May 2000 [20]. Releasing the experimental setup detail and result publicly made this stall-regulated wind turbine with a rated output power of 19.8 kW a popular case study for the validation of CFD-based studies and the development of new ideas [21, 22]. This stall-regulated turbine consists of two S809 airfoil-based blades, having a 10.058 m diameter and a thickness to chord ratio of 0.21. The airfoils are twisted in various sections (i.e., from $18.074°$ to $-1.816°$ at the root and tip) to improve aerodynamic performance in rotation time, specifically designed to be less sensitive to the surface roughness at the leading edge of the blade. Table 1 represents the blade characteristics [23]. The axial wind turbine is also shown in Fig. 2. We solely analyze the blades; thus, the tower is skipped.

**Table 1.** Specification of the NREL Phase VI wind turbine.

| | |
|---|---|
| Number of blade | 2 |
| Rotor radius (m) | 5.029 |
| Rotational speed (rpm) | 72 |
| Rated power (kW) | 19.8 |
| Rotational direction | CW |
| Blade tip pitch angle (deg) | 3 |

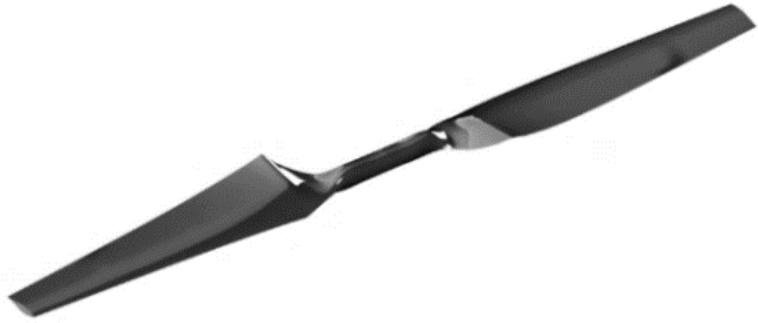

**Fig. 2**. Shape model of the NREL Phase VI wind turbine blade.

### 2.1. Governing equations and numerical schemes

The flow around the rotating axial wind turbine is simulated through three-dimensional unsteady RANS approach. The continuity and momentum equations along with the transport equations for the turbulent parameters are solved for the fluid phase. While an Eulerian-Lagrangian approach (i.e. DPM) deals with the flow/dust-particle interactions and neglects the inter-collisions of particles, in the case of dusty airflow. The dispersed phase exchanges the momentum with and is tracked through the fluid phase. The Reynolds-averaged continuity and momentum equations of a three-dimensional incompressible viscous flow (with constant properties) in the Cartesian coordinate system are

$$\frac{\partial U_i}{\partial x_i} = 0, \quad \text{and} \tag{1}$$

$$\rho\left(\frac{\partial U_i}{\partial t} + \frac{\partial U_i U_j}{\partial x_j}\right) = -\frac{\partial P}{\partial x_i} + \frac{\partial}{\partial x_j}(\mu \frac{\partial U_i}{\partial x_j} - \rho \overline{u'_i u'_j}) + \rho g_i + F_i \qquad (2)$$

where $U_i$ and $u'_i$ are the time-averaged and fluctuating velocity components, $\mu$ and $\rho$ represent the fluid viscosity and density and $F_i$ stands the force of interaction with the dispersed phase. The Reynolds stresses ($\rho \overline{u'_i u'_j}$) are related to the mean velocity gradients with the eddy viscosity ($\mu_t$) as follows:

$$-\rho \overline{u'_i u'_j} = \left[\mu_t(\frac{\partial U_i}{\partial x_j} + \frac{\partial U_j}{\partial x_i})\right] - \frac{2}{3}\rho k \delta_{ij}, \qquad (3)$$

in which $k$ is the turbulent kinetic energy.

To consider the turbulent nature of the flow properly, the shear stress transport SST $k$-$\omega$ model is selected. This consideration comes from a comparison of different turbulence models against experimental data [16, 23] and success of this model (i.e. the best among other turbulence models in following the wind tunnel results) in the previous attempts [24-27]. This two-equation model takes the advantages of the $k$-$\omega$ model in the near-wall region and the $k$-$\varepsilon$ model in the outer region, transforming by a blending function. Besides, SST $k$-$\omega$ incorporates some modifications, like considering a damped cross-diffusion derivative term in the $\omega$ equation, using different modeling constants and modifying the definition of the turbulent viscosity, compared with the standard $k$-$\omega$ model; thereby, it provides more reliable results for a wider class of flows such as airfoils.

This model includes two extra transport equations, i.e. turbulent kinetic energy ($k$) and the specific dissipation rate ($\omega$), to close the equations and consequently represent the turbulent properties of the flow (Eqs. 4 and 5).

$$\frac{\partial}{\partial t}(\rho k) + \frac{\partial}{\partial x_i}(\rho U_i k) = \Gamma_k + G_k - Y_k + S_k \qquad (4)$$

$$\frac{\partial}{\partial t}(\rho \omega) + \frac{\partial}{\partial x_i}(\rho U_i \omega) = \Gamma_\omega + G_\omega - Y_\omega + D_\omega + S_\omega \qquad (5)$$

In these equations, $G$, $\Gamma$, and $Y$ terms represent production, diffusion, and dissipation of turbulent kinetic energy ($k$) and the specific dissipation rate ($\omega$), respectively. Also, $D_\omega$ is the cross-diffusion term, popping up due to transforming the $k$-$\varepsilon$ model into $k$ and $\omega$ equations. $S_k$ and $S_\omega$ stand for the extra source terms. Explanations about how to calculate different terms and also the values of model constants can be found in [28].

In a Lagrangian reference frame, DPM integrates the force balance on the particle to track it. This force balance equates the particle inertia with the forces acting on the particle, and eventually form the equations of motions for the airborne particles (see Eqs. 6 and 7).

$$\frac{d\vec{u}_p}{dt} = \underbrace{\frac{3\mu C_D Re_r}{4\rho_p d_p^2}(\vec{u}-\vec{u}_p)}_{\gamma} + \underbrace{\frac{\vec{g}(\rho_p-\rho)}{\rho_p}}_{\psi} + \underbrace{\frac{1}{2}\frac{\rho}{\rho_p}\frac{d}{d_t}(\vec{u}-\vec{u}_p)}_{\xi} \quad (6)$$

$$\frac{d\vec{x}_p}{dt} = \vec{u}_p \quad (7)$$

The $\gamma$, $\psi$ and $\xi$ are the drag, gravity and added mass forces per unit particle mass, respectively. The vector symbol in the Lagrangian approach discussion distinguish it from the Eulerian one. So that, $\vec{u}$ and $\vec{u}_p$ denote the fluid and particle velocity vectors, while $\vec{g}$ = (0, -g, 0) and $\vec{x}_p$ = ($x_p$, $y_p$, $z_p$) is the dust particle Cartesian position. Moreover, $\rho_p$ and $d_p$ represent the particle density and diameter. The $Re_r$ (= $\rho d_p |\vec{u}_p - \vec{u}|/\mu$) in Eq. 6 is the relative Reynolds number and $C_D$ is the particle drag coefficient, which is estimated by Eq. 8. The constants $a_1$, $a_2$ and $a_3$ are given by Morsi and Alexander [29], applying to smooth spherical particles over several ranges of $Re_r$.

$$C_D = a_1 + \frac{a_2}{Re} + \frac{a_3}{Re^2} \quad (8)$$

The ordinary differential equations (ODE) of particle motions in Eqs. (6, 7) are solved by a 5$^{th}$-order Runge-Kutta scheme. Please note that the momentum exchange appears as a sink term (i.e. $F_i$) in Eq. 2.

*2.2. Boundary conditions and numerical algorithm*

The computational domain and the boundary conditions are shown schematically in Fig. 3. The computational domain has a radius of 8 rotor radiuses and the turbine is placed at distances of 4 rotor diameters from the inlet and 20 rotor diameters from the outlet. The domain is divided into two zones, stationary and rotational. A uniform velocity and a constant pressure are imposed at the inlet and outlet boundaries, respectively. Since the study aims to evaluate the CFD approach by comparing the RANS and wind tunnel results, the no-slip impermeable boundary condition is applied at the side wall as well as the surfaces of the blades, see also Sec. 3. The enhanced wall treatment method determines the rest conditions on the wall, where $k$ = 0 and a set of equations determines the boundary condition of specific dissipation ($\omega$) on the walls [30]. Although this method considerably improves the calculation of high gradients occurring near the walls, it needs a fine mesh with high resolution of $y^+$ < 1. The distribution of $y^+$ on the blade surface is depicted in Fig. 4.

The present work was carried out using Ansys Fluent. This finite-volume solver discretizes governing equations on a co-located grid arrangement. The Semi Implicit Method for Pressure-Linked Equations (SIMPLE) algorithm deals with the velocity-pressure coupling. The convective and diffusive terms are discretized with the second-order upwind and central differencing schemes, respectively.

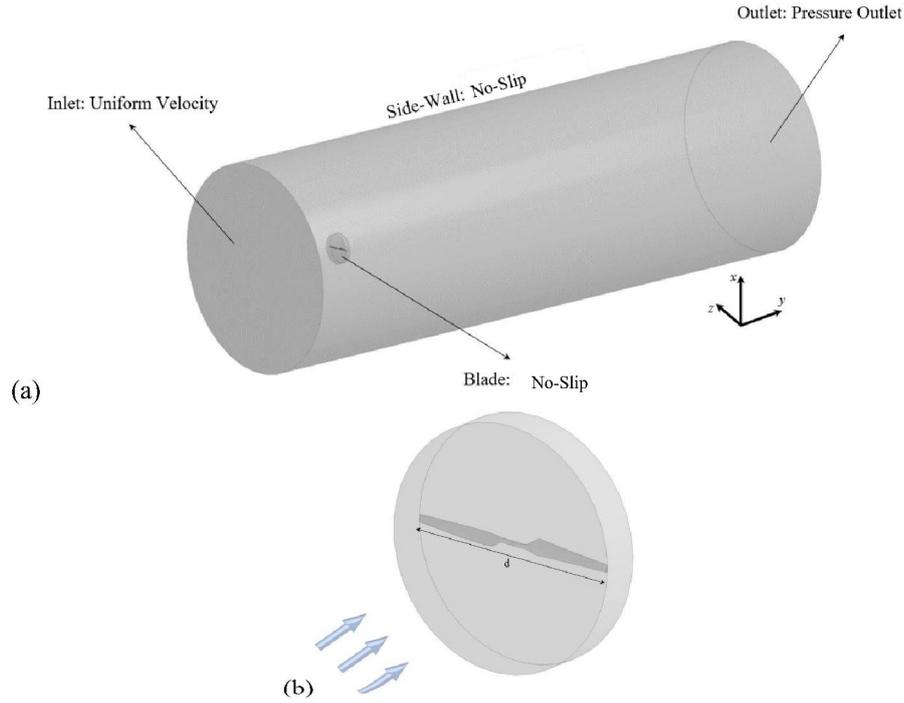

**Fig. 3.** Computational domain (not to scale) of the studied wind turbine including (a) stationary and (b) rotational zones.

For the clean conditions, the air flow field is calculated using the continuity and momentum equations along with the transport equations for the turbulent parameters. For dusty airflow conditions, the dust particles are distributed homogeneously in the inlet and injected with the same velocity as the free stream. Then, the equations of clean condition are solved first. Next, the dust particle trajectory is computed based on the air velocity field obtained in the previous step. Eventually to close the computational algorithm loop, the dust particle effect on the airflow is considered by adding the momentum source term into the continuous phase momentum equation. This two-way coupling procedure is continued at each time-step (0.001 s) to meet the convergence criterion of $10^{-6}$ for the scaled residuals.

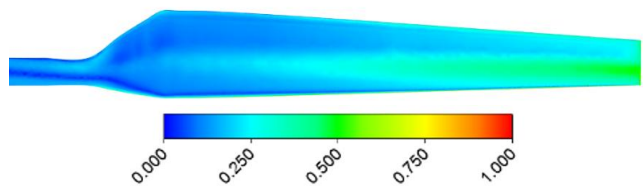

**Fig. 4**. The distribution of $y^+$ on the blade surface.

## 2.3. Grid generation

The structured prism elements for the boundary layer regions around the blades and unstructured tetrahedral elements for other regions are considered as computation mesh scheme for the wind turbine. The typical grids for the computational domain and blade are shown in Fig. 5 (a) and (b), respectively. Mesh refinements are defined at the rotational zone and near the blades to obtain flow details reliably. The *s* and *r* show the local coordinates stick to the blade.

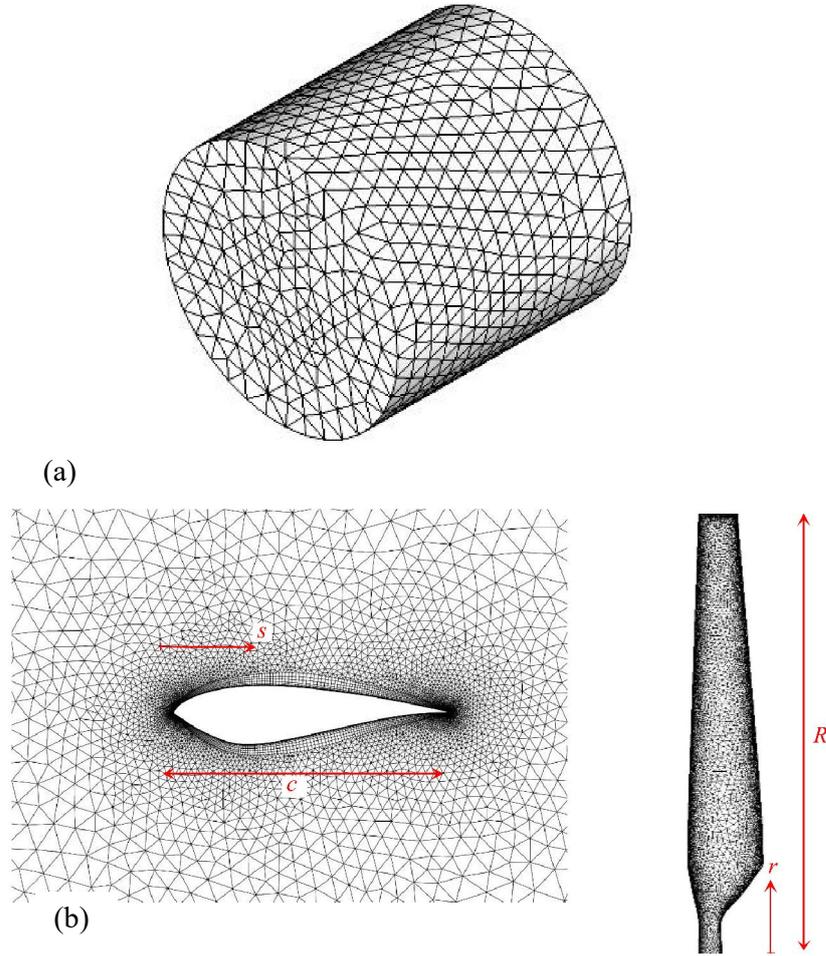

**Fig. 5.** Global view of the computational mesh for the (a) outer region and (b) blade. For the sake of clarity, a coarse grid is shown here although the pattern is similar to the fine ones.

The accuracy of spatial resolution was established by creating six computational grids. Table 2 compares the wind turbine power and thrust from the six grid systems for the clean air condition when $U_\infty = 7$ m/s and $AoA = 0°$. The time-averaged mechanical power (*Po*) and thrust ($T_r$) are obtained using Eq. 9. In this equation, *T* is torque, *A* is the cross-sectional area, $U_R$ is the wind velocity at the rotor, $U_\infty$ is the free stream velocity and $U_W$ is the wind velocity in the far wake of the rotor.

$$Po = T\Omega$$
$$T_r = \rho A U_R (U_\infty - U_W) \tag{9}$$

This table shows that increasing the number of cells from Mesh 4 (with 10357518 cells) to Meshes 5 and 6 causes a trifle change (i.e., less than 2%). Hence Mesh 4 is employed for all simulations in this study.

| Table 2. Grid independency study. | | | |
|---|---|---|---|
| Mesh | Number of cells | Power (KW) | Thrust (KN) |
| 1 | 3564891 | 7.222239 | 0.8572 |
| 2 | 5863252 | 4.931182 | 08375 |
| 3 | 7632549 | 5.699854 | 0.9235 |
| 4 | 10357518 | 5.915835 | 1.0520 |
| 5 | 11234518 | 5.925979 | 1.0522 |
| 6 | 12345518 | 5.920583 | 1.0523 |

## *2.4. Scope of parametric study*

Various conditions simulated in this work are listed in Table 3. A relatively large range of environmental and turbine operating conditions are taken into consideration. The dust particles in nature are usually in the range of 1-400 μm (i.e. up to 0.4 mm) [31]. The results, such as pressure coefficient distribution and force coefficients, are mostly presented for $d_p \leq 0.3$mm. It is also shown in Sec. 4 that the power and trust changes are saturated for $d_p \geq 0.3$mm. Nonetheless, the two largest $d_p$ in Table 3 are considered to study possible acute deterioration.

Although no similar study has been conducted so far, a few two-dimensional simulations studied the influence of dust particle size and volume fraction on the pressure distribution of airfoils, e.g., see [15]. The volume fraction ($Q$) is related to the diameter of ($d_p$) and rate of injected ($N$) particles by Eq. 10. The previous attempts observed a non-linear effects of $d_p$ and $N$ on the force and pressure coefficients for a given $Q$ (e.g. see [15, 19]). Although the individual interactions might be more significant in terms of momentum exchange between the continuum and discrete phases for larger $d_p$, fewer numbers of particles (i.e., $N$) at a fixed $Q$ can reverse the net dusty environment effect [15]. In other words, if the $N$ at a given $Q$ is reduced to study the influence of $d_p$, the $N$ effect on flow physics is dominated over $d_p$ at some level. For our study with a vast range of $d_p$ (see Table 3), a significant reduction of $N$ at the large $d_p$ values also considerably affect the results. That is why we here assume constant $N$ = 12000 for the simulations. Adjusting $Q$ with $\rho_p$, on the other hand, makes the analysis non-straightforward and complex (see Eq. 7). Thus, $\rho_p$ = 1900 kg/m³ for all simulations.

| Table 3. Various simulated conditions. | |
|---|---|
| $U_\infty$ (m/s) | 5, 7, 10, 13, 15, 20, 25 |
| $d_p$ (mm) | 0.025, 0.05, 0.1, 0.3, 0.6, 0.9 |
| $AoA$ (deg) | 0, 3, 9, 18.5, 25.5, 29.6, 37.8, 44 |

$$Q = \frac{\dot{m}_{dust}}{\dot{m}_{air}}, \qquad \dot{m}_{dust} = \frac{\pi}{6}\rho_p N d_p^3 \qquad (10)$$

The above discussion reveals that $Q$ is a direct function of $d_p^3$. The objective of the present study is to find how the turbine performance varies with increasing $d_p$. In fact, the present study aims to see if the NREL Phase VI wind turbine, which was designed to be less sensitive to surface roughness, is also reliable for the pollutant areas. The results prove the performance of the NREL Phase VI wind turbine does not undergo harsh drop till $d_p \approx 0.1$mm (Fig. 6) and almost is saturated for $d_p \geq 0.3$mm. The details are presented in Secs. 3 and 4.

## 3. Validation: Comparison of turbine power and thrust against experimental data

In order to evaluate the accuracy of the numerical procedure, the results (Eq. 9) are compared with the available experimental ones under clean air condition [23]. To do so, the governing equations (1–3) are solved without any source term (i.e. $F_i$). The results of single-phase flow simulations are presented in Fig. 6. It is worth noting that although the NASA Ames wind tunnel test-section is rectangular (i.e. 24.4 m×36.6 m) [23], a cylindrical domain is used in the current simulations. Given the relatively small blockage ratio (i.e., $\beta = 0.25\pi d^2$/test-section area) considered in both experimental and numerical studies (smaller than 0.09), the outer domain shape theoretically affects the global and wake topology results negligibly [32, 33]. On the other hand, the cylindrical domain generally improves the simulations by removing unessential grids and leads to more orthogonal grids.

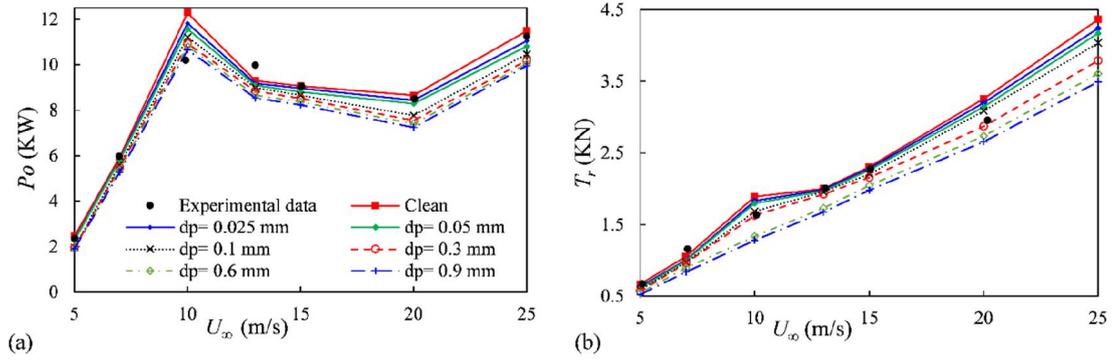

**Fig. 6.** Comparison of the simulated and experimental data [23]. (a) Power $Po$. (b) Thrust $T_r$.

Fig. 6 (a) indicates the variations of mechanical power against wind speed. Despite quantitative differences at $U_\infty = 10$ and 13 m/s, the trends of both measured and calculated data are similar. The power increases with the $U_\infty$ increment and reaches a maximum at $U_\infty = 10$ m/s and after a sharp decrement till $U_\infty = 13$ m/s hits a plateau from $U_\infty = 15$ m/s to 20 m/s. This behavior is attributed to the stall-regulated design of the turbine. Thus, the flow characteristics are related to the operation state of the turbine (i.e. pre-stall or post-stall zones). Except for $U_\infty = 10$ and 13 m/s, the CFD results show less than a 2% deviation from those of experiments. Maximum and minimum absolute deviations of 19.61% and 0.12% from experimental results are respectively observed at $U_\infty = 10$ m/s and 15 m/s, the average absolute deviation is also 5.14%. Fig. 6 (b) represents the turbine thrust ($T_r$) versus the wind speed. The thrust starts to increase with the $U_\infty$ increment due to its direct dependence to the wind speed (Eq. 9). An over-prediction of 15.7% is discernible at $U_\infty = 10$ m/s and the average absolute deviation from experimental results is 6.83%. The above comparison reveals that there is an acceptable agreement between the numerical and experimental results, although an over-prediction for $U_\infty = 10$ m/s cautions a higher uncertainty of results at this freestream velocity.

The surface streamlines on the suction side of the blade are presented for four wind speeds of 5, 7, 10 and 15 m/s (Fig. 7) to correlate the $Po$ and $T_r$ variations in Fig. 6 to the pre-stall and post-stall flow features over the blades. Five span-wise sections of r/R = 30%, 46.7%, 63.3%, 80% and 95% are also specified on the blade for better understanding of the descriptions. No flow separation is seen at $U_\infty = 5$ m/s and the streamlines largely signify the two-dimensionality of flow along the blade span, except near the hub region (up to r/R = 30%). The flow separates at the mid-chord when $U_\infty = 7$ m/s (Fig. 7). On the contrary of $U_\infty = 5$ m/s, a radial flow, apparently originating from the hub side, is observed after the flow separation line, where the reverse flow get in touch with the blade. The radial flow forms under the effect of the centrifugal acceleration and pressure gradient in this direction. Comparing streamlines at $U_\infty = 5$, 7, and 10 m/s evokes that the radial flow has much lower momentum compared to the mainstream, such that its footprint on the blade is observed only after the mainstream separation and where the reverse flow shows up. Flows at $U_\infty = 10$ and 15 m/s are highly three-dimensional and the hub flow radially spreads toward the tip. However, the nearly 2D separating flow at mid-chord (similar to what is seen for at $U_\infty = 7$ m/s and r/R > 30%) is observed more or less for r/R > 80% at $U_\infty = 10$ m/s. The separation occurs over the entire span at $U_\infty = 15$ m/s and the blade is fully stalled. Given the above observations, the authors believe that the higher uncertainty at $U_\infty = 10$ m/s is likely associated with the deficiencies of the RANS turbulence modelling in capturing highly transient effects.

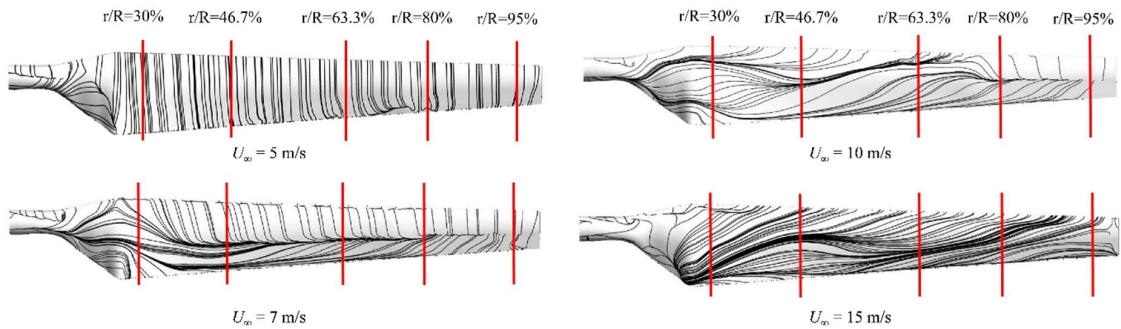

**Fig. 7.** Surface streamlines on the suction side of blade for different wind speeds and $AoA = 0°$.

## 4. Results and discussion

How the wind turbine performance is contingent on the dusty atmospheric environment is scrutinized here. A relative wide range of the airborne particle size ($d_p$), free-stream wind speed ($U_\infty$), and angle of attack ($AoA$) are studied (Table 3). First, the influences of $d_p$ and $U_\infty$ on the surface pressure distribution at $AoA = 0°$ are detailed. Then, the general dependence of the dusty wind physics over turbine on $AoA$ is surveyed by seeing drag and lift force values against $AoA$. Following this, the flow field on the blade, the turbine rotor wake, and the blade boundary layer specifications are discussed. Eventually, the variation of turbulent viscosity with $d_p$ is concisely investigated, as the effective parameter for the accumulation of dust on the blade.

### *4.1. Pressure distribution over blades*

Fig. 6 shows the influence of the particles on trends of power and thrust against the wind speed. A particle diameter ($d_p$) increase plummets the *Po* and $T_r$ down such that $d_p = 0.05$ and 0.9 mm reduce (the average for different wind velocities) the power 4.3% and 13.3% and thrust 4.3% and 20.5%, respectively. The dust has less impact on *Po* and $T_r$ at low wind speeds (i.e. up to $U_\infty = 7$ m/s), but the post-stall flows are highly contingent on the quality of oncoming freestream flow. From Fig. 6, it seems $d_p = 0.1$ mm (or 100 μm) is a threshold, at (for $U_\infty \geq 10$ m/s) and above which the $Po$ and $T_r$ drop sharply. Please note that the dusty wind in nature contains different ranges of particle sizes, and the $d_p$ here can be deemed as average. Besides, the oncoming wind is usually unsteady in the real world and the present study is a simplified case by removing those irregularities to fundamentally study the dependence of the wind turbine performance. The changes in the flow regime, and subsequently the blade pressure distribution, are likely the main reasons for the above variations. The latter is discussed here, while the former is presented concisely in the next sections.

Fig. 8 demonstrates measured [23] and computed pressure coefficient distribution at different span-wise sections (30%, 46.7%, 80% and 95%, Fig. 7) for $U_\infty = 5$ m/s and $AoA = 0º$. The time-averaged pressure coefficient is defined based on Eq. 11, where $p$ and $p_\infty$ represent local time-averaged pressure and reference pressure, respectively, and $\Omega$ (rad/s) is the angular velocity (Table 1). The typical pressure distribution ($p$) on the blade also has been also shown in Fig. 8. The $C_p$ is generally negative for the suction side and positive for the pressure side and the difference between the suction and pressure sides is maximum at the leading edge and almost constant after mid-chord (i.e. $s/c > 0.5$). The $C_p$ on the pressure side and right after leading edge decreases and then increases; it becomes negative at $s/c \approx 30\%$ and then positive at $s/c \approx 60\%$ (Fig. 8). These variations are the expected consequence of the underneath camber of the S809 airfoil (Fig. 5b). Whereas, the streamlines indicate the attached flow regardless of $s/c$ and r/R (not shown, but similar to Fig. 7 for $U_\infty = 5$ m/s).

$$C_p = \frac{p - p_\infty}{0.5\rho\left(U_\infty^2 + (\Omega r)^2\right)} \tag{11}$$

The numerical results closely follow experimental data for clean-air, giving more evidence on the essential validation for the subsequent analyses. However, small deviations are distinguishable on the suction side near the $s/c = 0$ which is more noticeable for r/R = 95%. The three-dimensionality of flow due to the tip vortices is likely responsible for the mentioned deviations. Similarly, Fig. 9 represents the measured [23] and computed $C_p$, but for $U_\infty = 10$, 15 and 25 m/s and different $d_p$ values with $AoA = 0º$. For $U_\infty = 10$ m/s knowing as the onset of stall [20], slight deviations from the experimental data are seen at different sections especially at the inboard zones (r/R= 30% and 46.7%). This echoes at $U_\infty = 15$ m/s as well, but mostly at the hub region (r/R= 30%). As pointed out in the previous section, these discrepancies can be explained by the radial flows and numerical difficulties from the transient stall effects. In addition, the sophisticated hub geometry and its instrumentation in the experiments introduce some deviations compared with simulations.

For $U_\infty = 15$ m/s (known as stall region), the flow becomes more complex and highly three-dimensional over the entire blade. The pressure difference between the suction and pressure sides is also increased along the blade but not near the tip region. The flow is separated over entire span for $U_\infty = 25$ m/s (deep stall region) and the pressure difference is increased considerably. It is shown later that increasing $U_\infty$ while having the same $AoA$ and blade rotational speed leads to facing the relative flow at the pressure side of the blade, i.e., a higher relative angle of attack. Therefore, the entire nominally blade suction side undergoes the reverse flow. It explains the negative relatively-invariant $C_p$ occurring in most r/R sections when $U_\infty$ = 15 and 25 m/s (i.e. post-stall).

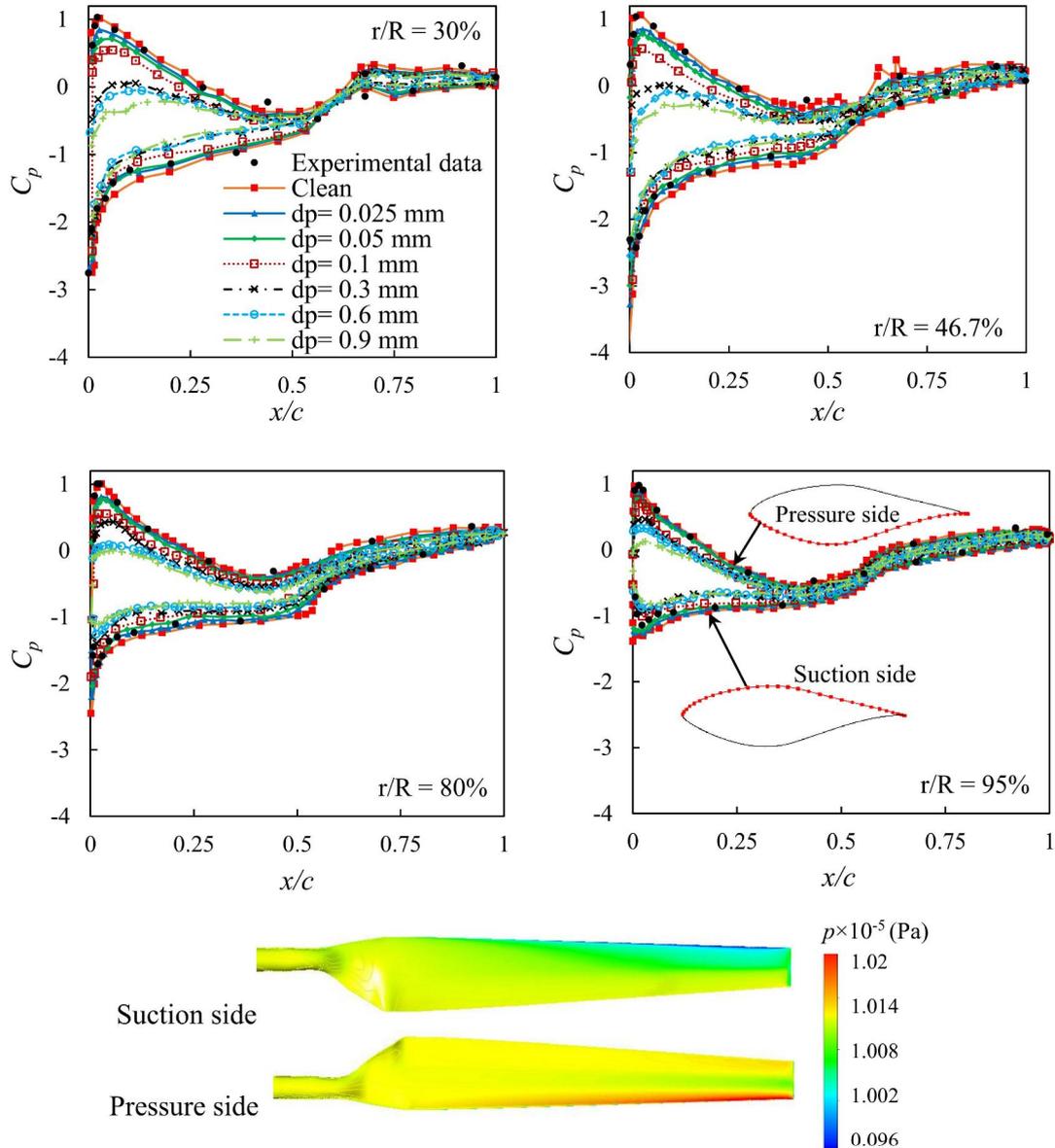

**Fig. 8.** Comparisons of the measured [23] and computed chord-wise pressure coefficient at span-wise sections of 30%, 46.7%, 80% and 95% for $U_\infty$ = 5 m/s and $AoA$ = 0°. The contours show the typical pressure distribution on the blade, here $d_p$ = 0 (clean).

The influences of adding dust particles with different $d_p$ on the $C_p$ are also portrayed in the Figs. 8 and 9. Irrespective to $d_p$ and $U_\infty$, the $C_p$ distributions largely resemble that of clean environment but the pressure difference between the both sides of the airfoil is decreased. Fig. 8 shows that the larger $d_p$, the lower of flow momentum. Furthermore, a higher $C_p$ drop is seen at the leading edge and on the adverse-pressure gradient region of the blade pressure side. It is also interesting from Fig. 8 that the dust with $d_p \geq 0.3$ puts off the flow system such that the $C_p$ peak on the blade pressure side slightly shifts downstream. Such shifting is also seen in Fig. 9 by increasing $U_\infty$, not because of the dust effect but, as discussed before, a higher effective angle of attack. Another important observation from Fig. 8 could be the subtle influence of $d_p$ until 0.1mm, intense between $d_p$ = 0.1mm to 0.3mm, and almost uniform again at $d_p \geq 0.3$mm. Similar observations are made at higher $U_\infty$ values, although Fig. 9 only presents $d_p \leq 0.1$mm to keep the clarity. However, the dust effect is less noticeable for the deep post-stall operating conditions ($U_\infty$ = 25 m/s) than

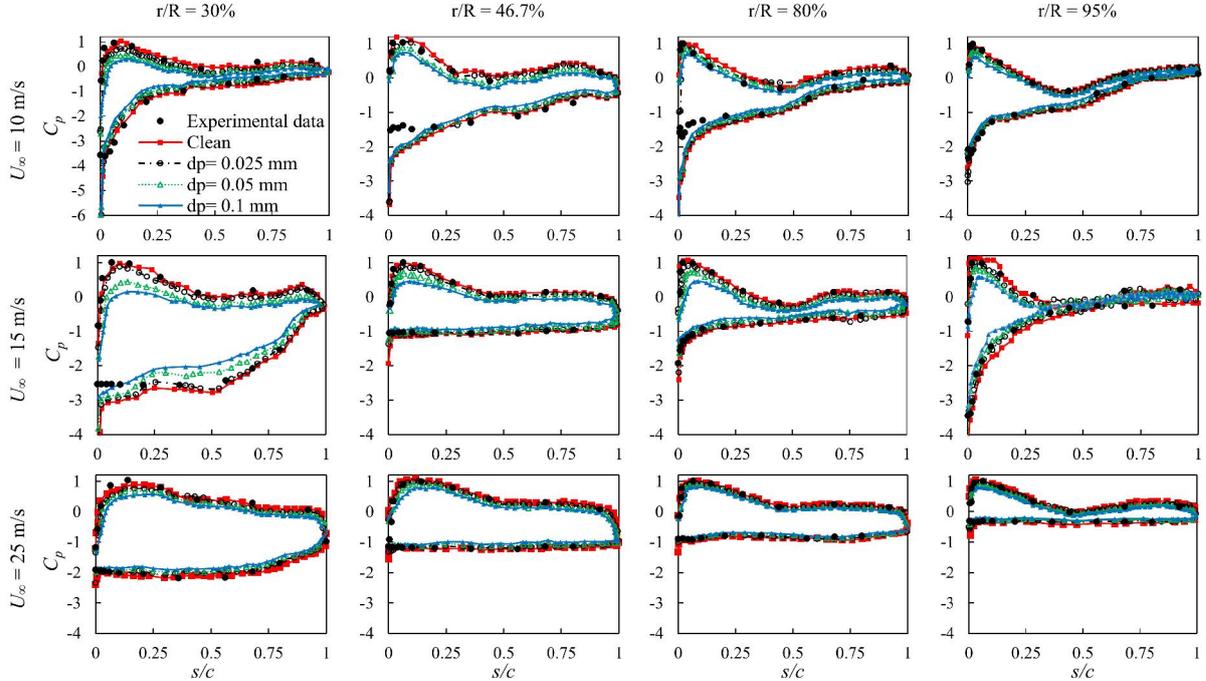

**Fig. 9**. Comparisons of the measured [23] and computed chord-wise pressure coefficient at span-wise sections of 30%, 46.7%, 80% and 95% for wind speeds of 10, 15 and 25 m/s and $AoA = 0°$.

### 4.2. Lift and drag forces versus blade angle of attack

The $C_p$ variation with $d_p$ and $U_\infty$ was studied in the previous section. But, the $AoA$ influence is yet to be clarified. The time-averaged sectional lift ($C_l$) and drag ($C_d$) coefficients, calculated by integrating $C_p$ (Eq. 11) in $x$ and $y$ directions, versus $AoA$ and $d_p$ are depicted in Figs. 10 and 11 ($U_\infty$ = 7 m/s). Increasing the $AoA$ makes the airfoil blunt and the blade suction side may fully or partially undergo the wake of separated flow. Also, the tip vortex strength and the radial flow pumping from the hub likely change with $AoA$ (not discussed here) and lead to increasing flow complexity. These effects cause the non-uniform variations of $C_l$ and $C_d$ along the blade span and complex variation of $C_l$ and $C_d$ as a function of $AoA$ in different sections (Figs. 10 and 11). For $AoA < 18°$, the $C_l$ generally increases along the span. But, the local values are smaller at both ends (i.e. r/R = 30% and 95%). For $AoA > 18°$, the $C_l$ shows different variations with $AoA$ along the span, e.g., it steadily increases up to $AoA = 30°$ at r/R = 30%, but largely decreases beyond $AoA = 20°$ at r/R = 46.7%. The $C_l$ is indeed affected reversely by moving toward the blade tip. This effect of tip vortex on recovering the base pressure has been documented for different end-free configurations in the literature (e.g. see [34, 35]). In general, the early half of the blade (i.e., r/R < 50%) bears higher $C_l$ values (Fig. 10). The influence of airborne particles on the lift is a downward translation and consequently aerodynamic performance degradation. Nonetheless, the value of reduction is not tangible for $d_p$ < 0.1mm. Introducing particles with larger $d_p$ amplifies this effect, but the $C_l$ is almost not sensitive to $d_p$ at smaller $AoA$, say $AoA \leq 10°$.

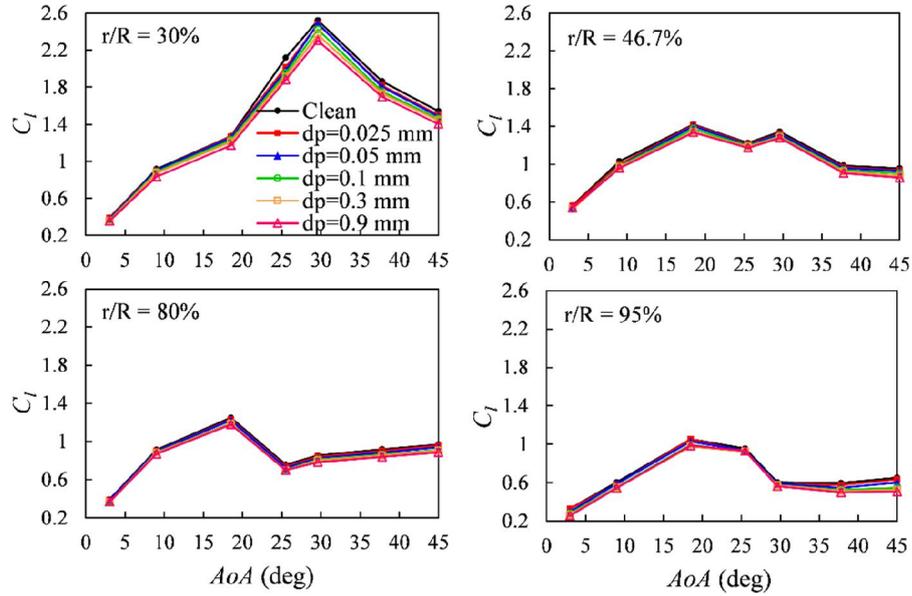

**Fig. 10.** The variations of lift coefficient against different angle of attacks at span-wise sections of 30%, 46.7%, 80% and 95% for wind speed of 7 m/s.

The $C_d$ variation with $AoA$ in Fig. 11 accords with the physical intuition and knowledge from the airfoils and wind turbines; the larger the $AoA$, the larger the $C_d$. In other words, increasing $AoA$ makes the blade blunter and the pressure drag rises significantly (Fig. 11). The above discussion on the effect of tip vortex on the base pressure and the three-dimensionality due to both spanwise-directed tip and radial flows explains the non-uniformity of $C_d$ along the blade span as well. Fig. 11 shows a larger $C_d$ in inboard sections (r/R < 50%). It means a larger $C_l$ is accompanied by a larger $C_d$ (Figs. 10, 11). The dust impact on the $C_d$ is also not profound, especially for $d_p$ < 0.1mm. This increase is less significant for smaller $AoA$ and $d_p$ values.

Despite some non-uniform variations of $C_l$ and $C_d$, Figs. 10 and 11 reveal the influence of $d_p$ on the flow is almost not affected by $AoA$. The simultaneous effect of $AoA$ and $d_p$ was also examined at $U_\infty$ = 15 m/s. But it is not presented owing to the similarity of observations with the above. Consequently, the focus is only given to $AoA = 0°$ in the next sections, where the dusty wind influence on the turbine flow physics is sought.

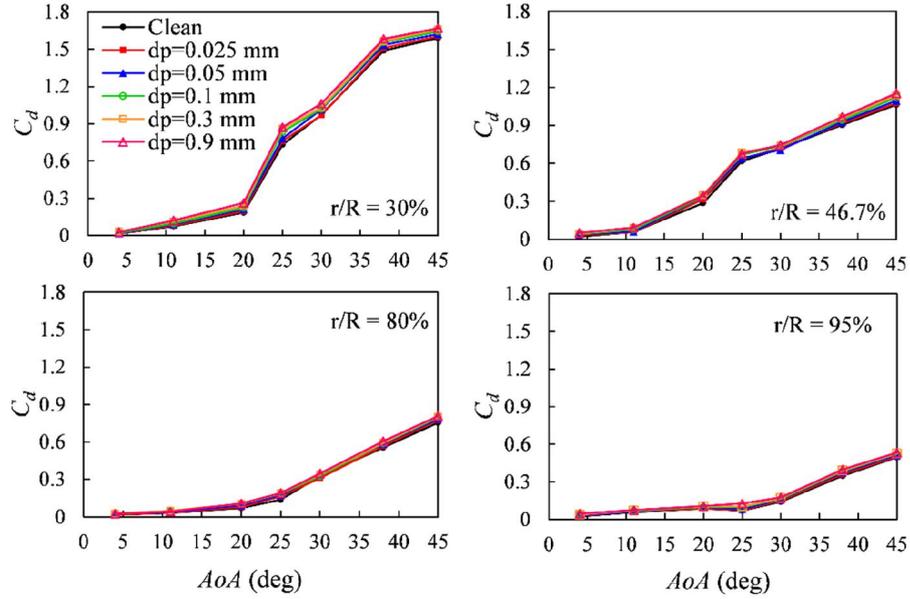

**Fig. 11.** The variations of drag coefficient against different angle of attacks at span-wise sections of 30%, 46.7%, 80% and 95% for wind speed of 7 m/s.

*4.3. Wake flow*

Figs. 12-14 represent the velocity contours and the streamlines over the blade at r/R = 30%, 46.7%, and 95% for the clean and dusty winds, respectively. The results approve that the flow is largely attached at $U_\infty = 5$ m/s along the blade span and $d_p$ has no considerable effect on the flow regime at r/R = 30%, 46.7% (Figs. 12, 13). Nonetheless, a reduction of the velocity on the leading edge in response to the momentum exchanges between dust particles and the main flow is more conspicuous at r/R = 95% (Fig. 14), perhaps due to a higher relative velocity magnitude. As a result, the flow is separated on the tail of the upper camber when $d_p = 0.3$ and 0.9 mm. The smaller pressure at r/R = 95% was also observed in Fig. 8. This momentum loss, inducing the $C_p$ change, reduces $C_l$ and $C_d$ subsequently (see Sec. 4.1).

Due to the relative velocity between the rotational blade motion and oncoming free-stream flow, the flow is considerably distorted over the blade at $U_\infty > 5$ m/s (Figs. 12-14). It leads to a more uniform $C_p$ on the blade pressure side at higher $U_\infty$ values (Fig. 9). For $U_\infty = 10$ m/s under clean wind condition (Figs. 12-14), the separation point shifts to the leading edge at r/R = 30% and 46.7%, and the flow is attached only to a small region of the blade tip (r/R = 95%) (see also Fig. 7). The boundary layer separation and flow recirculation are also the cases for $U_\infty > 10$ m/s, irrespective of the $d_p$ (Figs. 12-14). Similar to contours for $U_\infty = 5$ m/s, a significant velocity reduction over the airfoil is evident for the post-stall cases as well ($U_\infty \geq 10$ m/s), which again betokens the momentum loss due to the dust (Figs. 12-14). Regardless of the $U_\infty$, the reverse flow over the blade suction side shows two bubbles. A $U_\infty$ increase gradually accommodates the whole upper face in the wake. This copes the variation of $C_p$ in Figs. 8 and 9 on this surface and shows why the $C_p$ is almost uniform on the suction side at $U_\infty = 15$ and 25 m/s (see also Fig. 7). Adding particles into the oncoming flow results in smaller velocity in the wake, which is a signature of stronger reverse flow [36, 37]. That is, the dust advances the boundary layer separation (see also Sec. 4.4) and the flow recirculation becomes more severe. This effect is amplified by increasing $d_p$.

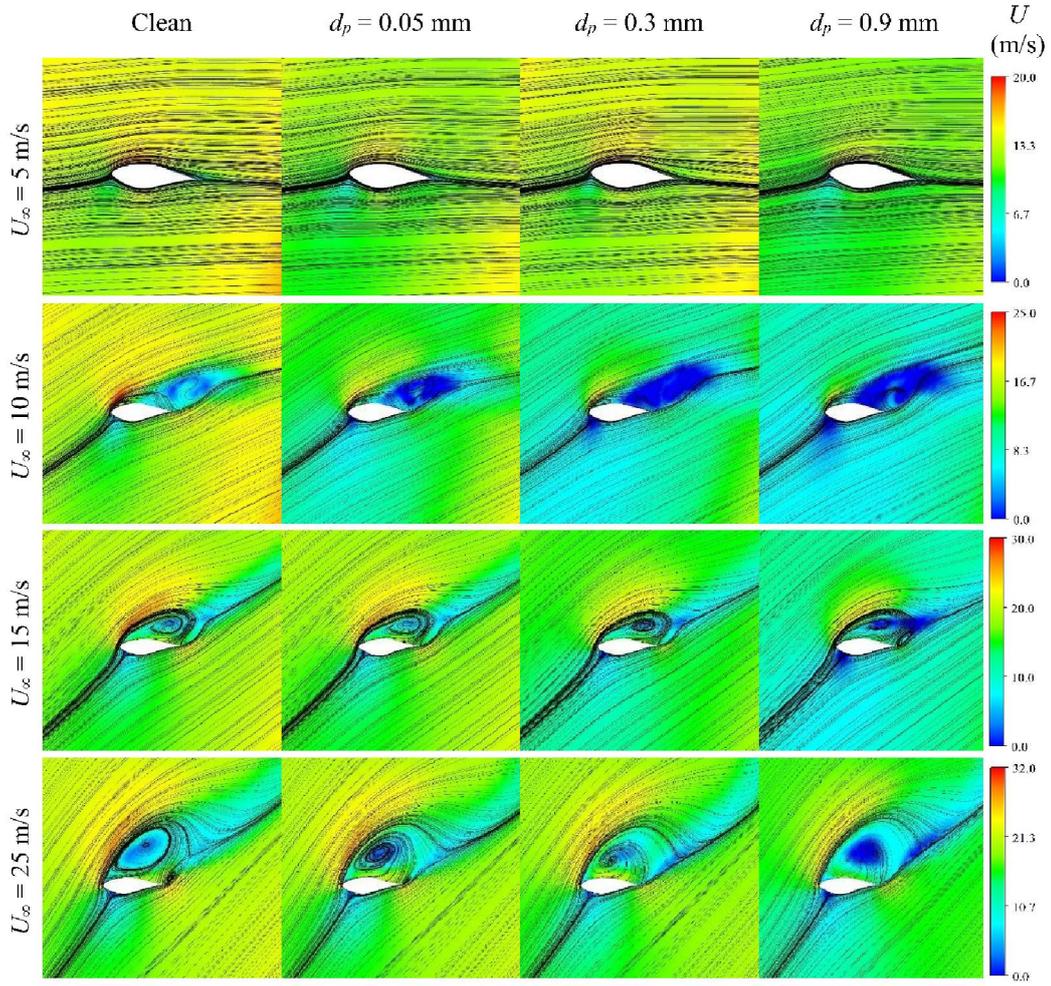

**Fig. 12.** Velocity contour superimposed with flow streamlines for different wind speeds and particle diameters at r/R = 30%. (*AoA* = 0°)

The time-averaged normalized axial velocity ($U_y$) profiles for $U_\infty$ = 7 and 25 m/s and $d_p$ = 0.05, 0.3 and 0.9 mm are plotted at upstream ($y/d$ = -1) and different downstream ($2 \leq y/d \leq 20$) locations in Figs. 15 and 16. The velocity profiles have been normalized by $U_\infty$. The $U_\infty$ = 7 m/s for the clean wind is the representative of the attached flow in the pre-stall condition, while $U_\infty$ = 25 m/s is the separated and post-stall ones. Owing to the no-slip condition on the outer boundary, resembling the wind tunnel condition, the boundary layer develops on this wall. The upstream profiles at $y/d$ = -1 shows that the oncoming flow is almost independent of $d_p$. While the profiles differ significantly in the wake. This observation indicates that the momentum exchange becomes dominant in our simulations when the flow interacts the turbine and also in the turbulent flow motion. The wind turbine extracts the air momentum, resulting in a velocity deficit and forming a W-shaped velocity profile in the near-wake. This shape can be attributed to the aerodynamic design of the blades and the rotor hub. Also, the wake characteristics behind the turbine are largely dependent to the variation of angle of attack on the turbine blade because of its direct influence on the variation of axial and angular momentum of a wind turbine [20]. The velocity deficit is intensifies with increasing $d_p$. But, Fig. 6 shows a smaller extracted power for larger $d_p$. The larger velocity deficit for the dusty wind might be because of a different flow state on the blade and the momentum exchange between the particles and flows in the turbulent wake. For instance, it was shown in Fig. 12-14 (but for $U_\infty$ = 5 m/s) that the flow separation in the blade tip region comes to being for the dusty wind, while it is not the case when the wind is clean.

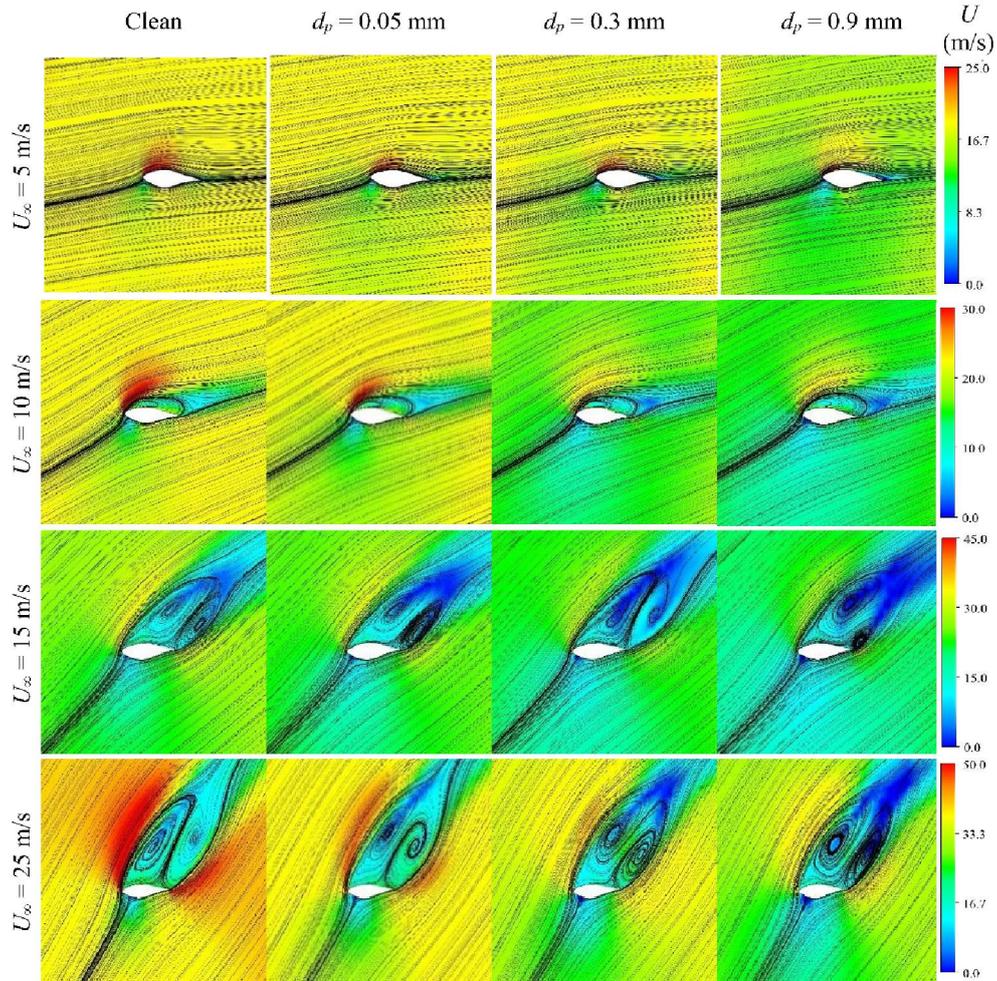

**Fig. 13.** Velocity contour superimposed with flow streamlines for different wind speeds and particle diameters at r/R = 46.7%. ($AoA = 0^o$)

The redistribution of the momentum from outside into the wake through turbulence recovers the velocity deficit in the far-wake. In fact, transferring the momentum into the wake causes the wake to expand and velocity deficit reduction. The peaks of the W-shaped profiles which correspond to the lowest velocity deficits occur in the inboard regions of the wind turbine. After $y/d$ =12 the W-shaped profiles turn into the V-shaped ones. On the other hand, the $y$ range corresponding to the velocity deficits, where $U_y/U_\infty = 0.99$, becomes larger at first by moving downstream and then this trend is reversed, showing the wake width variation.

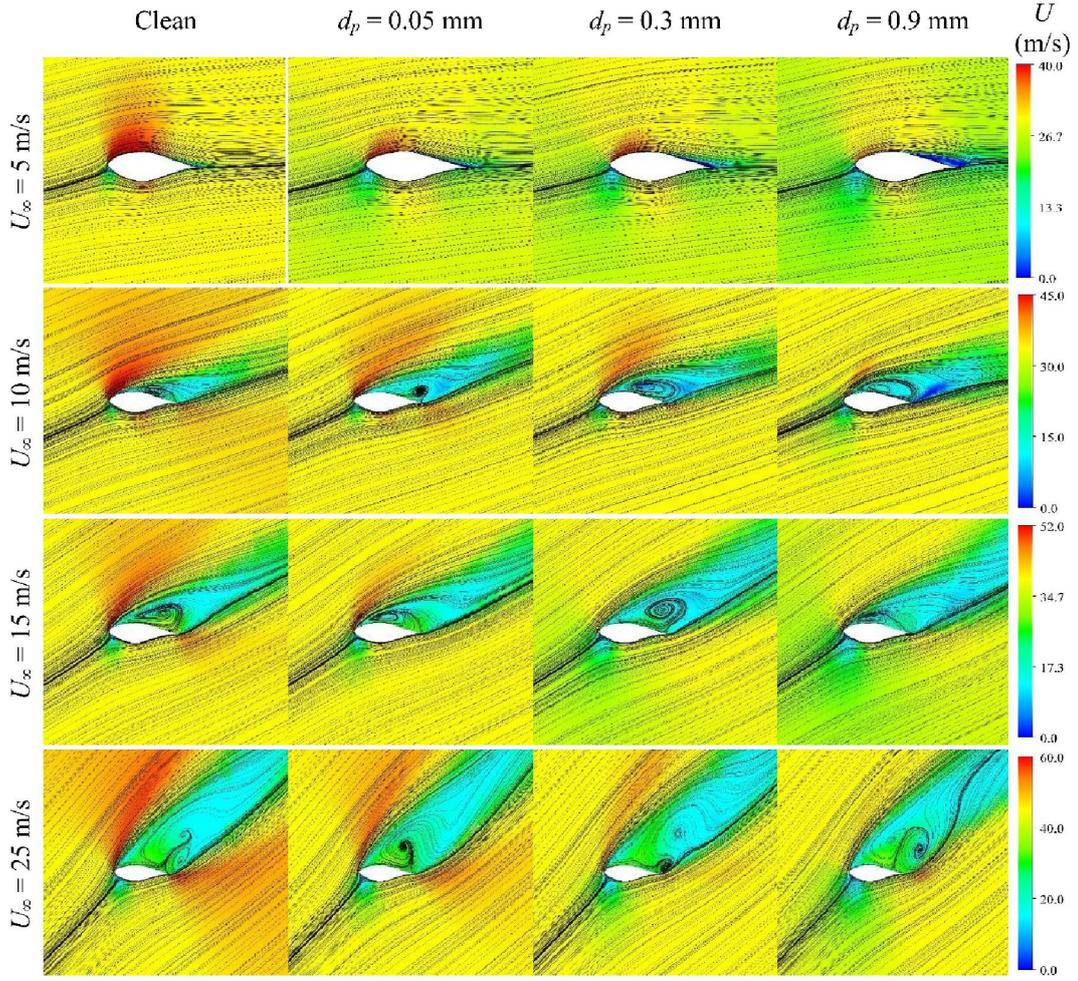

**Fig. 14.** Velocity contour superimposed with flow streamlines for different wind speeds and particle diameters at r/R = 95%. (*AoA* = 0º)

Although the above observations are largely akin in Figs. 15 and 16, the wake width and velocity deficits are contingent on $U_\infty$. Figs. 15 and 16 reveal that the maximum momentum extraction and hence the normalized velocities are dependent on the free-stream wind speed and particle diameter in the regions corresponding to the turbine blades. The velocity deficits are more pronounced at $U_\infty$ = 7 m/s and large $d_p$ values than those of $U_\infty$ = 25 m/s. This does not imply more power generation, as the *Po* is a function of the $U_\infty^3$, but signifies higher percentages of the air momentum extraction at $U_\infty$ = 7 m/s. In fact, the flow is attached to the entire blade surface at $U_\infty$ = 7 m/s (compared with $U_\infty$ = 25 m/s) and consequently a larger proportion of energy is extracted from the passing flow. Given the fact that the velocity profiles of clean wind are analogous to dusty, it may be concluded that the role of particles is to reduce the momentum of the main flow in the turbulent wake as well. Increasing $d_p$ smoothens the jagged velocity profiles at $y/d \geq 8$ (Figs. 15 and 16), likely showing the far-wake instabilities. While temporal averaging was performed for a similar time range. The momentum exchange through particles leads to this damping effect.

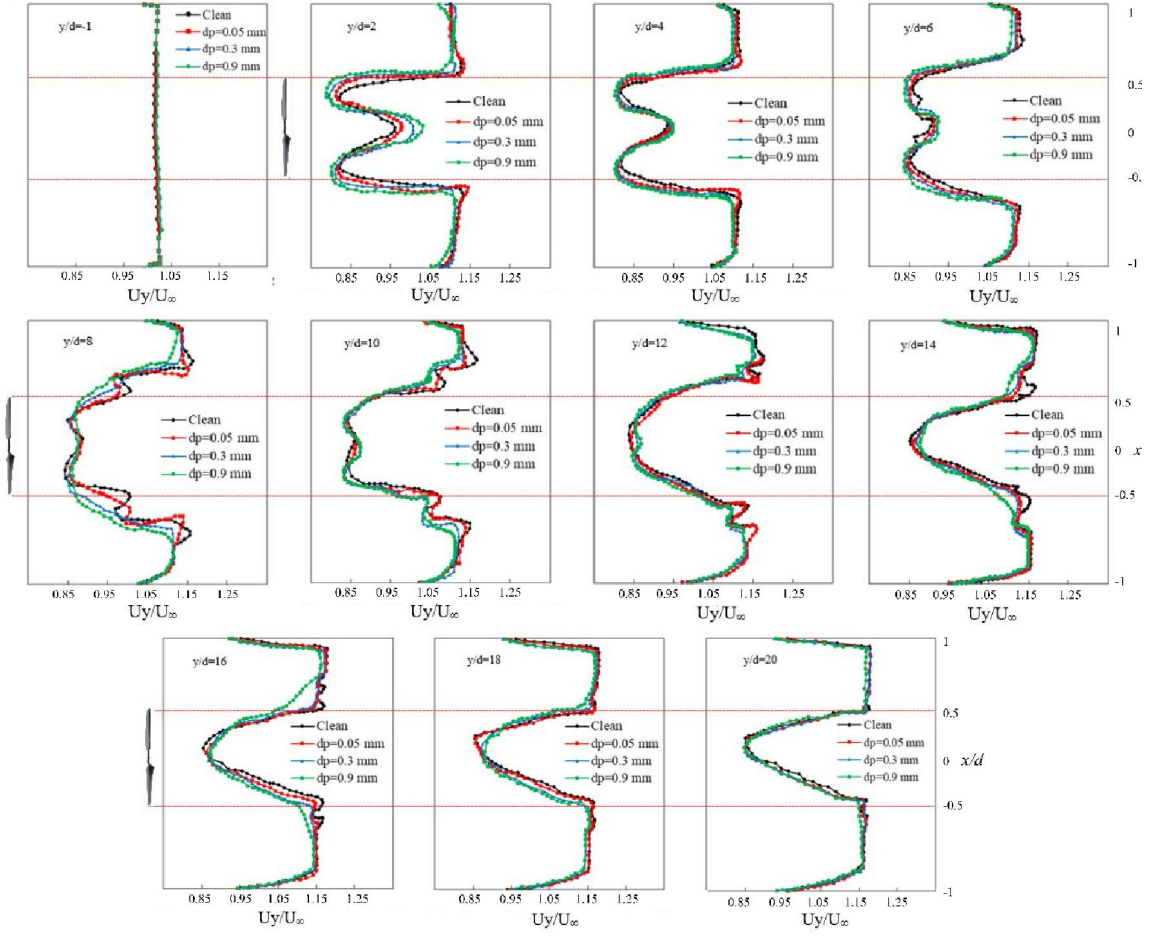

**Fig. 15.** Comparison of the normalized y-velocity profiles for different upstream and downstream locations ($U_\infty = 7$ m/s and $AoA = 0°$). The red dotted lines show the blades' extent in the domain.

*4.4. Boundary layer separation*

It was shown in the previous section that the airborne particles reduces the momentum of the continuum phase and subsequently change the flow field over the blades. Advancing the boundary layer separation and strengthening the reverse flow were determined as the main impacts. Then, it was observed that the velocity deficits are intensified in the near-wake and the far-wake instabilities are dampened with $d_p$ although the time-mean wake flow largely keeps its general format. On the other hand, it was shown in Sec. 4.1 that $Po$ and $T_r$ undergo remarkable changes for larger $d_p$ values. The following section tries to answer why the $C_p$ undergoes a drastic reduction at $d_p \geq 0.3$ and how advancing the flow separation plays role in this regard. This goal is chased by portraying the boundary layer velocity profiles at four chord-wise positions on the upper surface (suction side) of the blade in Figs. 17-19, where r/R = 30%, 46.7%, and 95%, and $U_\infty$ = 5, 10, and 15 m/s are considered.

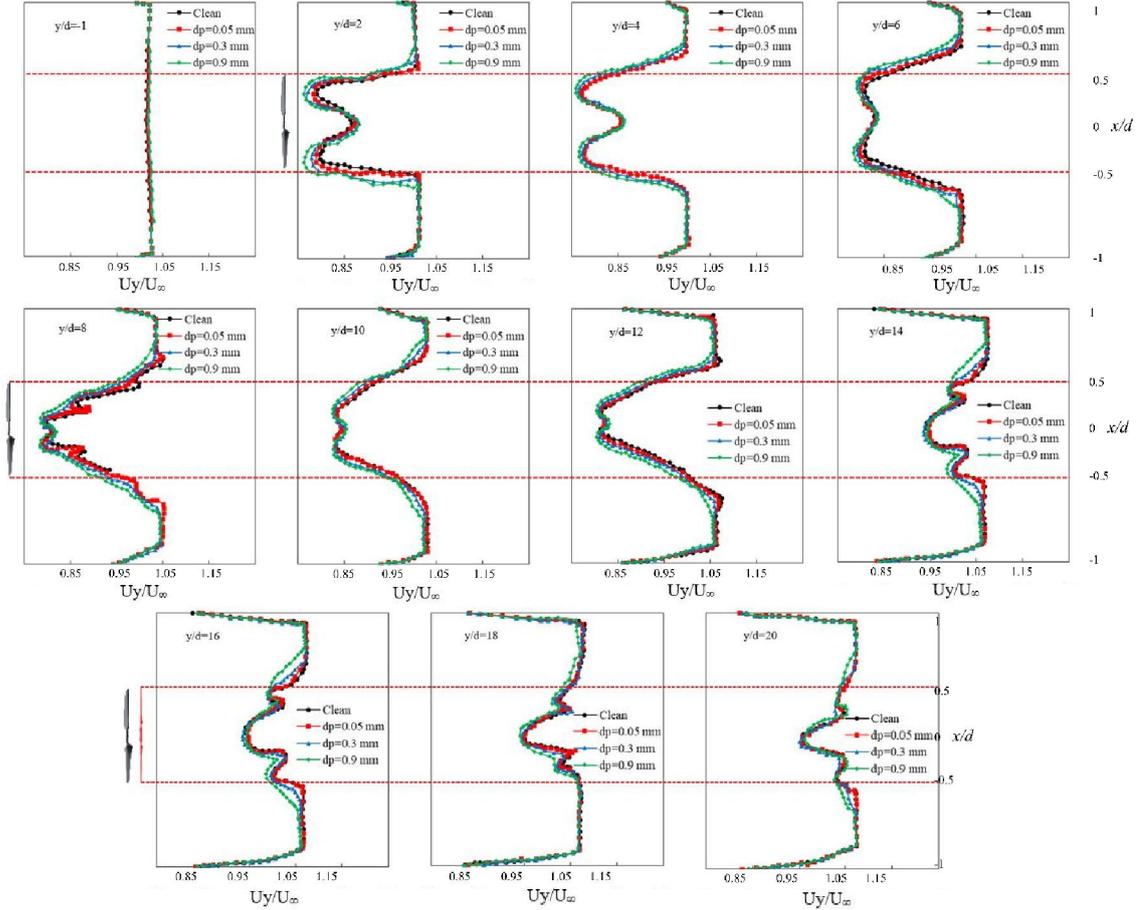

**Fig. 16.** Comparison of the normalized y-velocity profiles for different upstream and downstream locations ($U_\infty$ = 25 m/s and $AoA$ = 0°). The red dotted lines show the blades' extent in the domain.

Fig. 17 indicates changing the boundary layer behavior from $s/c \geq 0.5$ due to turning the favorable pressure gradient over the blade to the adverse one. Despite this transformation, the flow does not separate when $d_p$ = 0 and 0.05 mm. The boundary layer for the winds with $d_p$ = 0.3 and 0.9 mm, however, shows a kink in the profile, which implies the flow separation. This observation follows those from the velocity contours in Figs. 12-14 and pressure coefficients in Figs. 8 and 9. The reason for the above fact is also seen in Fig. 17 by noting that irrespective of the measurement location on the blade, introducing the dust particles into the main flow decelerates the boundary layer; the larger the particles, the more severe the decline. Given the observations in Secs. 4.1 and 4.3, one can conclude that the airborne particles with an average $d_p \geq 0.3$ are supercritical in terms of the continuum phase momentum reduction. So, it leads to flow separation and considerable $Po$ and $T_r$ reduction (Fig. 6).

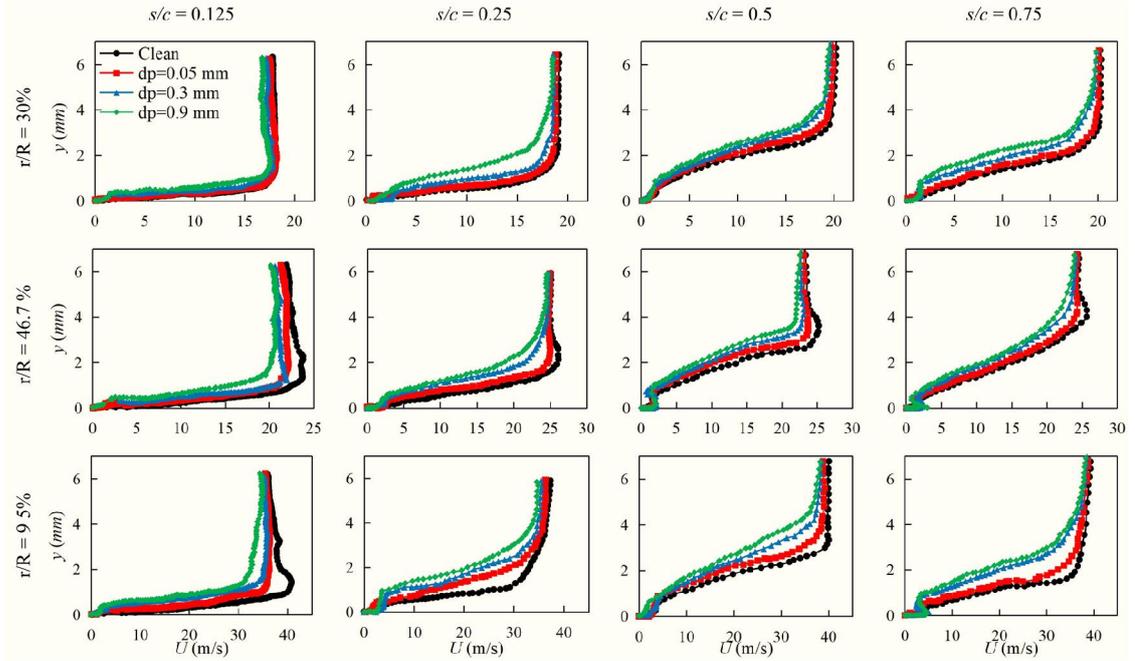

**Fig. 17.** Boundary layer velocity profiles at four chord-wise and three span positions on the upper surface of the turbine blade at $U_\infty = 5$ m/s and $AoA = 0°$. The $y$ indicates the distant from wall and perpendicular to the $s$-direction in Fig. 4.

Although Fig. 17 explains the $d_p$ impact on the flow physics of particle-wind-turbine interactions, the rate of $Po$ and $T_r$ reduction is much higher at $U_\infty \geq 10$ m/s (Fig. 6). In other words, the degree of the $d_p$ influence on the flow physics is a function of $U_\infty$. For $U_\infty = 10$ and 15 m/s (Fig. 18), the dusty winds with $d_p \geq 0.3$ separate even at $s/c = 0.125$ and the kink in the boundary layer velocity profiles gradually moves to the larger $y$. This shows the recirculation bubble is expanded steadily with an $s/c$ increment. A comparison of Figs. 17-19 indicates that the airborne particles have more profound effect on the deviation of the velocity profile of dusty winds from the clean ones at the post-stall states. That is, in addition to the de-energization of the boundary layer and the flow separation almost from the leading edge, the momentum exchange inside the reverse flow above the blade is another momentous effect of particles. This explains the higher rate of $Po$ and $T_r$ reduction for the pre- and post-stall states (Fig. 6).

One more point that is worth noting is the size of the recirculation bubble over the blade. The bubble size, in the sense discussed in the previous paragraph, is larger at r/R = 46.7% than that of r/R = 30% and 95% (Figs. 18 and 19). Looking back into the previous sections shows us that the highly three-dimensional flow at the hub and tip regions shrinks the recirculation bubble on both ends of blade. This observation proves the necessity of a three-dimensional simulation here and reflects why the 2D simulations of turbine sections usually cannot illustrate the flow physics thoroughly or estimate the global quantities properly.

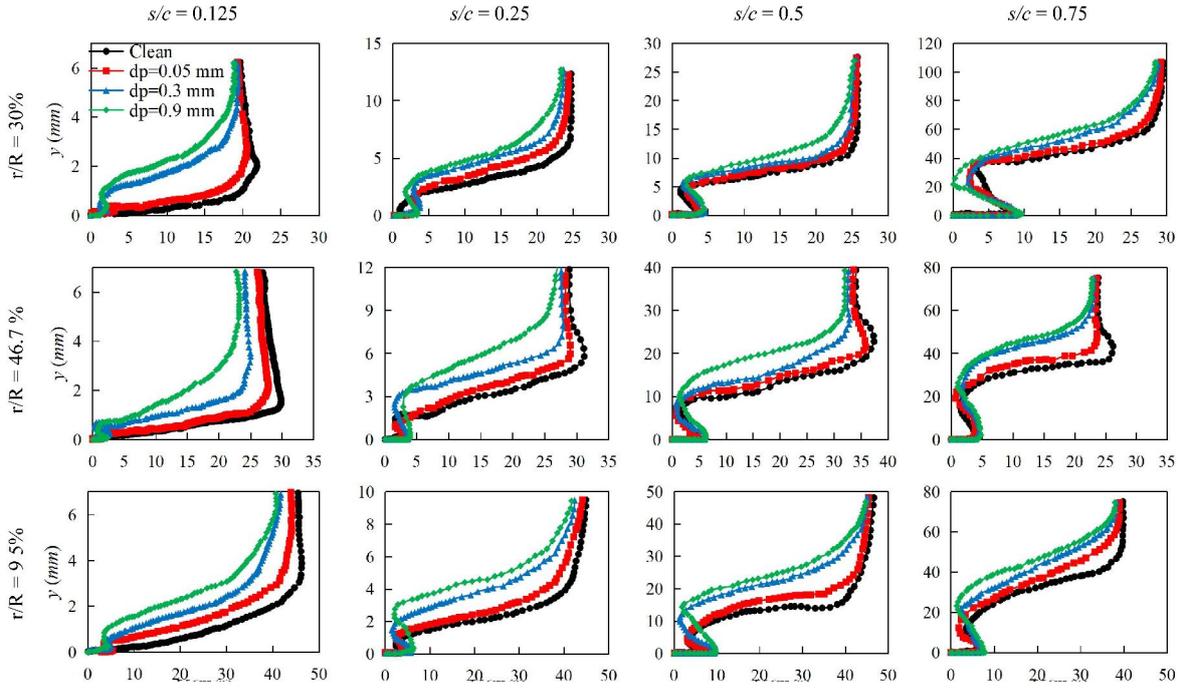

**Fig. 18.** Boundary layer velocity profiles at four chord-wise and three span positions on the upper surface of the turbine blade at $U_\infty = 10$ m/s and $AoA = 0°$. The $y$ indicates the distant from wall and perpendicular to the $s$-direction in Fig. 4.

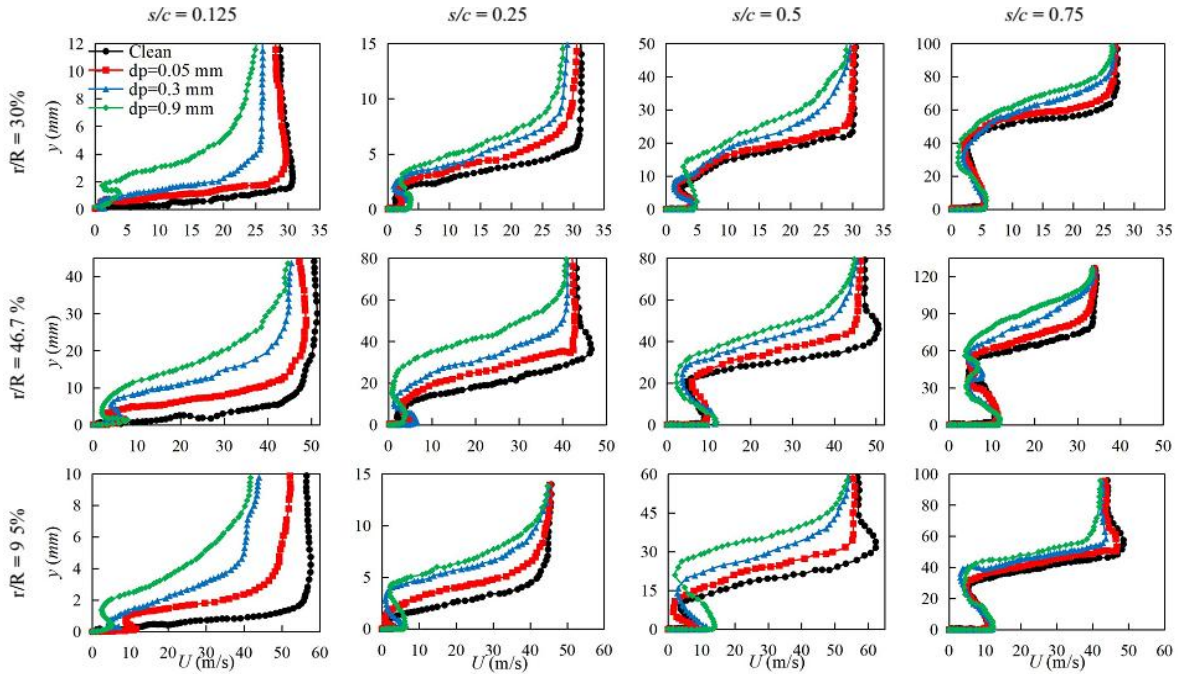

**Fig. 19.** Boundary layer velocity profiles at four chord-wise and three span positions on the upper surface of the turbine blade at $U_\infty = 15$ m/s and $AoA = 0°$. The $y$ indicates the distance from wall and perpendicular to the $s$-direction in Fig. 4.

## 4.5. Turbulent viscosity

Some studies in the literature considered the dust effects as blade roughness due to accumulation. To link the present study to that point of view, this section concisely discusses the variation of turbulent viscosity in the boundary layer. The turbulent viscosity is a leading parameter of particle accretion [5, 15]. Similarly, Wu et al. [5] studied the turbulent viscosity variation over an axial wind turbine under different rainfall rate. It was found that the heavier rain considerably increases the turbulent viscosity, which subsequently considered a source of water accumulation on the blade and larger skin friction drag. Fig. 20 presents the influence of dust particles on turbulent viscosity ($\mu_t$) in the boundary layer for $U_\infty$ = 5 and 15 m/s. This figure gives the mean spatiotemporal $\mu_t$ over the wind turbine blade. Increasing $d_p$ causes rising the dust volumetric percentage in the flow. It is observed from Fig. 20 that the $\mu_t$ in the present study also increases with $d_p$ in the whole domain and consequently in the boundary layer. The particles are not clustered on the blade in our simulations. Changing the velocity gradient very next to the blade (Figs. 17-19) is deemed the main source of skin friction. The contribution of this factor with the pressure change due to the particles (Figs. 8 and 9) leads to considerable drag variation (Fig. 11). However, Fig. 20 shows that an increase of $d_p$ and/or $U_\infty$ enhances the $\mu_t$ in the boundary layer; thus, the probability of dust accretion would increase with this theory.

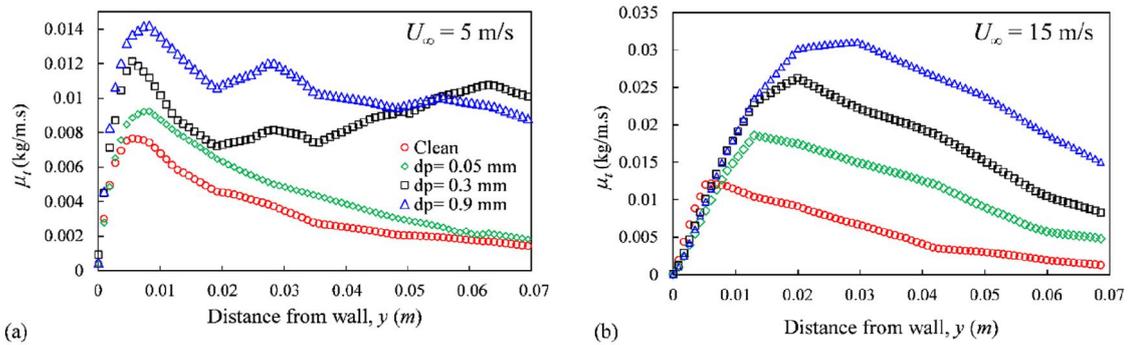

**Fig. 20.** The mean temporal-spatial turbulent viscosity (a) over the blade and (b) in the boundary layer.

## 5. Conclusion

Due to the importance of ambient atmospheric environment on performance of wind turbines, a NREL Phase VI wind turbine under dusty air conditions was investigated using a 3D CFD approach. The study includes relatively large range of effective parameter, i.e., the wind speed $U_\infty$ = 5-25 m/s, particle diameter $d_p$ = 0.025-0.9 mm and angle of attack $AoA$ = 0°-44°. An Eulerian-Lagrangian multiphase approach modeled the dusty air flow over the turbine. The air as the primary phase was anticipated through solving the RANS equations associated with the SST $k$-$\omega$ turbulence model in an Eulerian reference frame and the dust particles as the second phase were treated using the equations of motion for particles in a Lagrangian reference frame. The two phases were coupled via a momentum source term in the continuous phase momentum equation. The mentioned model was validated against the wind tunnel data from literature for the clean air condition at first and then was applied to examine the influences of dusty air condition.

The results indicate deterioration of the turbine performance, more intense effects for the larger particle. The airborne particles with the average $d_p = 0.1$ mm were determined as a threshold, and an acute reduction of wind turbine power and thrust occurs for the higher values. It was also found that the variation of blade angle of attack on the aforementioned particle effects is moderate. The momentum exchange between the particles and continuum phase generally de-energises the airflow. The particle effects are decisive in advancing the boundary layer separation and strengthening the reverse flow over the suction side. The former plays a principal role in the pre-stall state, while both are momentous in the post-stall one. The time-mean wake flow is analogous for different $d_p$ values, but the velocity deficits are intensified in the near-wake, and instabilities in the far-wake are dampened with $d_p$. It was also discussed that the growth of $d_p$ and $U_\infty$ increases the turbulent viscosity in the boundary layer, as a key parameter in the occurrence of dust accretion.

The obtained results provide valuable information about the effects of a dusty environment on wind turbine performance. Besides, it gives some narrower criteria and standards for choosing suitable sites for new wind farms or optimizing the blades. Given the present results, the influence of dusty winds on wind farms can be a worthy object for future focus.